# The propagating mechanism of Chapman-Jouguet deflagration

Yunfeng Liu [a,b,*], Wenshuo Zhang [a,b], Zijian Zhang [a,b]

[a] *Institute of Mechanics, Chinese Academy of Sciences, No.15 Beisihuanxi Road, Beijing, 100190, China*

[b] *School of Engineering Science, University of Chinese Academy of Sciences, No.19(A) Yuquan Road, Shijingshan District, Beijing, 100049, China*

**Abstract**

The deflagration-to-detonation transition (DDT) process is of great importance to both combustion theory and industry safety. In this study, the propagating mechanism of Chapman-Jouguet (C-J) deflagration is studied. Firstly, three models are put forth to decouple the C-J detonation front. These three models are (a) to introduce an expansion parameter into the one-dimensional energy equation, (b) to increase the activation energy of the chemical reaction model and (c) to decouple the shock wave from the flame front by artificial method. The C-J deflagration is obtained after the C-J detonation is decoupled by one-dimensional numerical simulations with different models, chemical reaction kinetics and numerical schemes. Secondly, the propagating mechanism of C-J deflagration is discussed. For the C-J deflagration with a propagating velocity of about 1/2 C-J detonation, the static temperature behind the leading shock wave is too low to ignite the combustion. But, the total temperature of the flow induced by the leading shock wave is high enough to ignite the mixture. The induced flow is slowed down by the rarefaction waves form the wall and its static temperature increases. The flame and the leading shock wave propagate with almost the same velocity and the double-discontinuity structure of the flow field keeps stable. The propagating velocity equals to the sound speed of the combustion products, which is about 1/2 C-J detonation velocity.

# 1 Introduction

The study of deflagration-to-detonation transition (DDT) process is of great importance to the combustion theory and industry safety. It is one of the major unsolved problems in combustion and detonation theory. It is also an extremely interesting and difficult scientific problem because of the complex nonlinear interactions among the different contributing physical processes, such as turbulence, shock interactions and energy release.

______________________________________________________________________

The DDT process was first observed in experiments by Brinkley and Lewis [1]. Oppenheim and his coworkers did much work and had a very deep insight into DDT process [2-4]. They obtained very clear schlieren photos and found that shock-shock interaction plays an important role in DDT process. Oppenheim derived the locus of the thermo-states downstream of the Chapman-Jouguet deflagration (C-J deflagration) front on the hypothesis that the front is of the C-J type and called this particular locus a Q-curve [5-6].

Chue and Lee experimentally studied two types of quasi-steady high-speed deflagration [7]. One was the reaction-waves created in and propagating through rough tubes containing obstacles; the other was deflagrations created from established C-J detonations by eliminating the transverse waves from the triple-wave structure. Both classes of deflagration were observed to travel at about one half of the corresponding C-J detonation speed. Lee called this deflagration wave quasi-detonation or C-J deflagration.

Zhu and Lee experimentally used the reflection of a C-J detonation on a perforated plate to generate high speed deflagrations downstream in order to investigate the critical conditions that lead to the onset of detonation [8]. They found that the critical deflagration speed is not dependent on the turbulence characteristics of the perforated plate but rather on the energetics of the mixture like a C-J detonation. The eventual onset of detonation was postulated to be a result of the amplification of pressure waves that leads to the formation of local explosion centers via the Shock Wave Amplification by Coherent Energy Release (SWACER) mechanism during the pre-detonation period.

Saif et al. experimentally studied the fast flames and their transition to detonation for five different hydrocarbons [9-10]. Their experimental results indicated that these C-J deflagrations dynamically restructure and amplify into fewer stronger modes until the eventual transition to detonation. The transition length to a self-sustained detonation was found to correlate very well with the mixtures' sensitivity to temperature fluctuations. They constructed the normalized DDT length by the characteristic distance, which was reflective of the induction delays behind the leading shock wave.

Goodwin and Oran conducted multi-dimensional numerical simulations to study different interactions leading to the DDT process [11]. The configuration studied was a long rectangular channel with regularly spaced obstacles containing a stoichiometric mixture of ethylene and oxygen, initially at atmospheric conditions and ignited in a corner with a small flame. The initial laminar flame developed into a turbulent flame with the creation of shocks, shock-flame interactions, shock-boundary layer interactions and a host of fluid and chemical-fluid instabilities. The result may be eventually the deflagration-to-detonation transition.

In this paper, we studied the propagating mechanism of C-J deflagration by theoretical analysis and one-dimensional numerical simulations. In order to create the C-J deflagration by decoupling the C-J detonation front, we put forth three methods, (a) to introduce an expansion parameter in the one-dimensional energy equation, (b) to increase the activation energy of chemical reaction model and (c) to decouple the shock wave from flame front by artificial method. The C-J deflagration was obtained successfully by one-dimensional numerical simulations and the mechanism of C-J deflagration was discussed.

**2 Physical models**

The DDT process usually occurs under the critical condition where the propagating velocity of C-J deflagration is about 1/2 C-J detonation. We theoretically studied the DDT process and put forth a physical model [12]. The structure of the flow field during DDT process is illustrated in Fig.1a. The flow field is divided into three regions by the primary shock wave SW and the flame. Region 1 is the premixed detonable mixture at initial pressure and temperature and at rest. Region 2 is the preheated mixture behind the SW and region 3 is the combustion products. The secondary shock wave SW' is a shock wave produced by constant-volume combustion in region 2. It's a function of the leading shock wave and the flame.

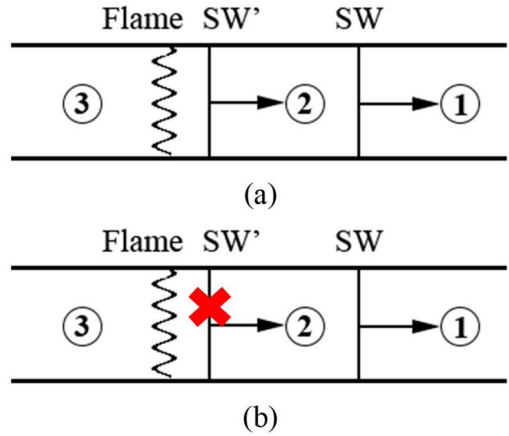

**Fig. 1** Structures of the flow field during DDT process (a) and the C-J deflagration wave (b) in the laboratory coordinates

Suppose there is a critical state that when the secondary shock wave SW' catches up and merges with the leading shock wave, the thermodynamic parameters of the new transmitted shock wave are exactly equal to the thermodynamics parameters of a C-J detonation, and therefore, a DDT process occurs. We obtained this critical DDT criteria in Eq. (1),

$$\frac{6M_1^2}{M_1^2+5}=\frac{7M_{C\text{-}J}^2}{6}\frac{T_1}{T_0} \tag{1}$$

where, $M_1$ is the Mach number of the leading shock wave under critical condition, $M_{C\text{-}J}$ is the Mach number of C-J detonation in the initial detonable mixture, $T_0$ is the total constant-volume combustion temperature and $T_1$ is the initial static temperature, respectively.

The C-J deflagration can also exit under this critical condition propagating with a constant velocity for a certain distance without the occurrence of DDT process. The structure of C-J deflagration is illustrated in Fig.1b. In this critical case, the combustion does not produce pressure rise and no secondary shock wave SW' is produced by combustion. As a result, no DDT process occurs. Therefore, the essence of C-J deflagration is constant-pressure combustion.

In this study, we study the propagating mechanism of C-J deflagration by one-dimensional numerical simulations and theoretical analysis. We create the C-J deflagration wave from C-J detonation by decoupling the detonation front. In order to decouple the detonation front, we put forth three different physical models which are presented as follows.

**(a) Introducing an expansion parameter in the one-dimensional energy equation**

In order to obtain the C-J deflagration, the first step is to decouple the shock wave from the flame front. The first physical model is by introducing an expansion parameter in the one-dimensional energy equation. This process is completed by simplifying the three-dimensional governing equations. The three-dimensional Euler equations for detonation simulation with one-step overall reaction model are as follows,

$$\frac{\partial \rho}{\partial t} + \frac{\partial \rho u}{\partial x} + \frac{\partial \rho v}{\partial y} + \frac{\partial \rho w}{\partial z} = 0 \tag{2}$$

$$\frac{\partial \rho u}{\partial t} + \frac{\partial \rho u^2 + p}{\partial x} + \frac{\partial \rho uv}{\partial y} + \frac{\partial \rho uw}{\partial z} = 0 \tag{3}$$

$$\frac{\partial \rho v}{\partial t} + \frac{\partial \rho uv}{\partial x} + \frac{\partial \rho v^2 + p}{\partial y} + \frac{\partial \rho vw}{\partial z} = 0 \tag{4}$$

$$\frac{\partial \rho w}{\partial t} + \frac{\partial \rho uw}{\partial x} + \frac{\partial \rho vw}{\partial y} + \frac{\partial \rho w^2 + p}{\partial z} = 0 \tag{5}$$

$$\frac{\partial \rho e}{\partial t} + \frac{\partial (\rho e + p)u}{\partial x} + \frac{\partial (\rho e + p)v}{\partial y} + \frac{\partial (\rho e + p)w}{\partial z} = 0 \tag{6}$$

$$\frac{\partial \rho Z}{\partial t} + \frac{\partial \rho u Z}{\partial x} + \frac{\partial \rho v Z}{\partial y} + \frac{\partial \rho w Z}{\partial z} = \dot{\omega} \tag{7}$$

$$e = \frac{RT}{\gamma - 1} + \frac{1}{2}(u^2 + v^2 + w^2) + Zq \tag{8}$$

$$\dot{\omega} = -k\rho Z \exp\left(-\frac{E_a}{RT}\right) \tag{9}$$

where, $\rho$ is the gas density, $u, v, w$ the velocity, $e$ the energy density, $Z$ the mass fraction of reactants, $q$ the density of heat release, $\gamma$ the specific heat ratio, $\dot{\omega}$ the mass formation rate of combustion products, $k$ the pre-exponential coefficient, $E_a$ the activation energy, $T$ the temperature, $R$ the gas constant, respectively.

Suppose the main flow direction for C-J detonation and C-J deflagration is in $x$ direction and the partial derivatives along the $y$ and $z$ directions can be ignored. In addition, we are only interested in the flow in $x$ direction and do not need to know the values of flow parameters in other directions. Therefore, the momentum equations in $y$ and $z$ directions can be deleted. The simplified governing equations are as follows,

$$\frac{\partial \rho u}{\partial t} + \frac{\partial \rho u^2 + p}{\partial x} = 0 \tag{10}$$

$$\frac{\partial \rho u}{\partial t} + \frac{\partial \rho u^2 + p}{\partial x} = 0 \tag{11}$$

$$\frac{\partial \rho e}{\partial t} + \frac{\partial (\rho e + p)u}{\partial x} = 0 \tag{12}$$

$$\frac{\partial \rho Z}{\partial t} + \frac{\partial \rho u Z}{\partial x} = \dot{\omega} \tag{13}$$

$$e = \frac{RT}{\gamma - 1} + \frac{1}{2}\left(u^2 + v^2 + w^2\right) + Zq \tag{14}$$

$$\dot{\omega} = -k\rho Z \exp\left(-\frac{E_a}{RT}\right) \tag{15}$$

The simplified governing equations are very similar to the one-dimensional inviscid governing equations. The only difference is that the energy equation of Eq.(14) contains the kinetic energy $v^2$ and $w^2$ in $y$ and $z$ directions, which cannot be ignored. Considering the kinetic energy in the main flow direction, the total kinetic energy of Eq. (14) can be expressed as,

$$\frac{1}{2}\left(u^2 + v^2 + w^2\right) = \frac{1}{2}Cu^2 \tag{16}$$

A new parameter $C$ is introduced in this simplified one-dimensional energy model, which represents the kinetic energy in all flow directions. In Eq. (16), the parameter $C \geq 1$ is physical, while $C < 1$ is unphysical. In the numerical simulations, for the finite-rate chemical reaction model, the flow in the detonation introduction zone could be a transonic flow. In this case, the parameter $C$ represents the expansion effects in three-dimensional space. The expansion effect decreases the static temperature to be below the auto-ignition temperature, which decouples the flame front from the shock wave and create the C-J deflagration.

**(b) Increasing the activation energy of chemical reaction models**

The second method to make the flame front and the shock wave decoupled of the C-J detonation is by increasing the activation energy $E_a$ of the chemical reaction model by increasing the parameter $\alpha$ of Eq.(17) in the traditional one-dimensional governing equation. By increasing the activation energy, the C-J detonation will become unstable and finally quench. In this model, the parameter $C$ is not needed.

$$\dot{\omega} = -k\rho Z \exp\left(-\frac{\alpha E_a}{RT}\right) \tag{17}$$

**(c) Decoupling the shock wave from flame front by artificial method**

The third method to make the flame front and shock wave decoupled of a C-J detonation is by closing and opening the chemical reaction in the program artificially. We know that for the one-dimensional C-J detonation, the maximum gas velocity is at the detonation front, so we choose the velocity $u$ as a parameter and define a threshold velocity $u^*$. The model is expressed in Eq. (18) as follows,

$$\text{chemical reation} = \begin{cases} \text{closed if } (u \geq u^*) \\ \text{open \quad if } (u < u^*) \end{cases} \tag{18}$$

## 3 Numerical methods

The computational domain is a one-dimensional tube with the left-end closed and the right-end open. The length of the tube is 1m. The premixed detonable mixture is $H_2$/Air with an equivalence ratio of ER=1.0. The initial pressure and initial temperature are 0.1MPa and 300K, respectively. The C-J detonation is initiated at the closed-end with a high-temperature and high-pressure region and propagates from left to right. The detailed parameters of the overall one-step model can be found in [13,14].

The governing equations are numerically solved by using a three-order WENO scheme. The flux vector is split by Steger-Warming splitting method and the time marching integration is performed using the three-order TVD Runge-Kutta integration. The uniform mesh is 0.1mm and the resolution study of 0.05mm is also conducted. The mirror reflection boundary condition is applied on the closed-end. A 9-species, 19-reactions detailed chemical reaction kinetic model [15,16] is also used to validate the conclusion. In the program of detailed chemical reaction kinetic model, the advection upstream splitting method (AUSM) scheme is applied to compute the convective fluxes.

## 4 Results and Discussion

In the one-dimensional numerical simulation, the C-J detonation is first initiated. Then, the C-J deflagration wave is created by adjusting the expansion parameter $C$, increasing the activation energy parameter $\alpha$ or the threshold velocity $u^*$. The C-J deflagration waves simulated by the above three models and the detailed chemical reaction kinetic are displayed in Fig.2a-d, respectively. We can find from Fig.2 that the C-J deflagration can be simulated successfully by any model and chemical reaction kinetics. This means that the one-dimensional numerical results are correct and physical. The propagating velocity of C-J deflagration is about 1/2 C-J detonation velocity. And the transition process from C-J detonation to C-J deflagration is abrupt.

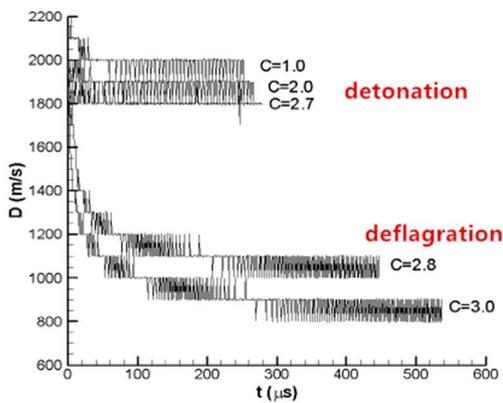 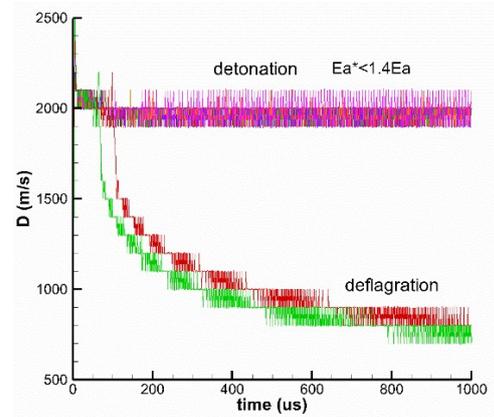

(a) model-a        (b) model-b

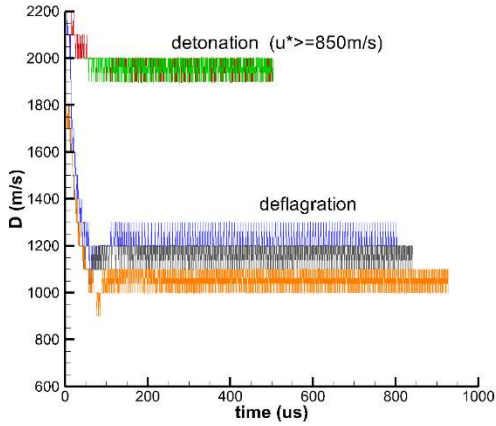
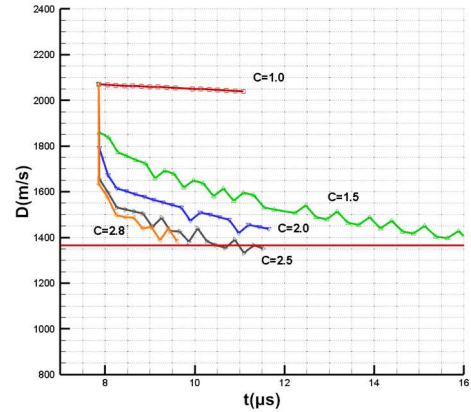

(c) model-c          (d) detailed chemical reaction kinetics with model-a

Fig.2 C-J detonation waves simulated by three different models

The pressure profiles of C-J detonation and C-J deflagration simulated by model (a) are plotted in Fig.3, respectively. From Fig.3a we can see that for $C \leq 2.7$, the C-J detonation is not be quenched. The pressure profiles are very similar. The C-J detonation front is followed by rarefaction waves and the point of zero-velocity is at the center point between the detonation front and the closed-end.

But the pressure profile of C-J deflagration in Fig.3b is completely different from that of C-J detonation. Firstly, the length of rarefaction region is very short, and the zero-velocity point is very close to the C-J deflagration front. Secondly, the pressure at and behind the deflagration front is much lower than that of C-J detonation. For instance, the pressure at C-J detonation front is about 2.5MPa, while the pressure at C-J deflagration front is about 0.9MPa, which is very close the pressure at the closed-end wall of about 0.6MPa. This means that C-J deflagration can be considered as constant-pressure combustion.

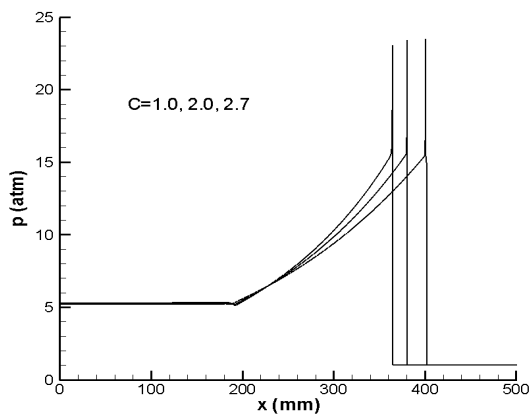
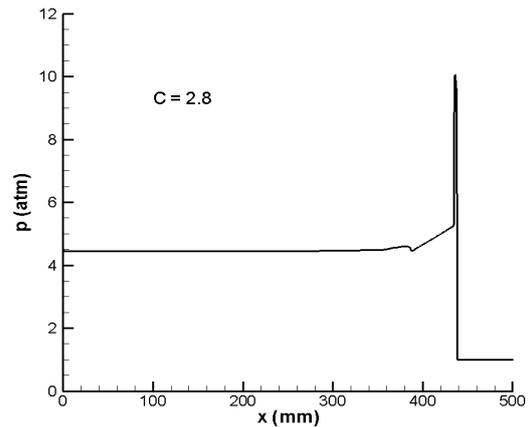

(a) C-J detonation          (b) C-J deflagration

Fig.3 Pressure profiles of C-J detonation and C-J deflagration

The profiles of different parameters of C-J deflagration are plotted in Fig.4. The parameters are static temperature, velocity, Mach number and chemical reaction parameter Z, respectively. The propagating mechanism of C-J deflagration can be explained as follows. For the C-J detonation, the shock wave and flame front are tightly coupled with each other. The combustion is initiated by the static temperature behind the shock wave. Once the shock wave and the flame are decoupled by the above three methods, the heat release cannot support the shock wave and the propagating velocity decays. As a result, the static temperature behind the shock wave is too low to ignite the combustion.

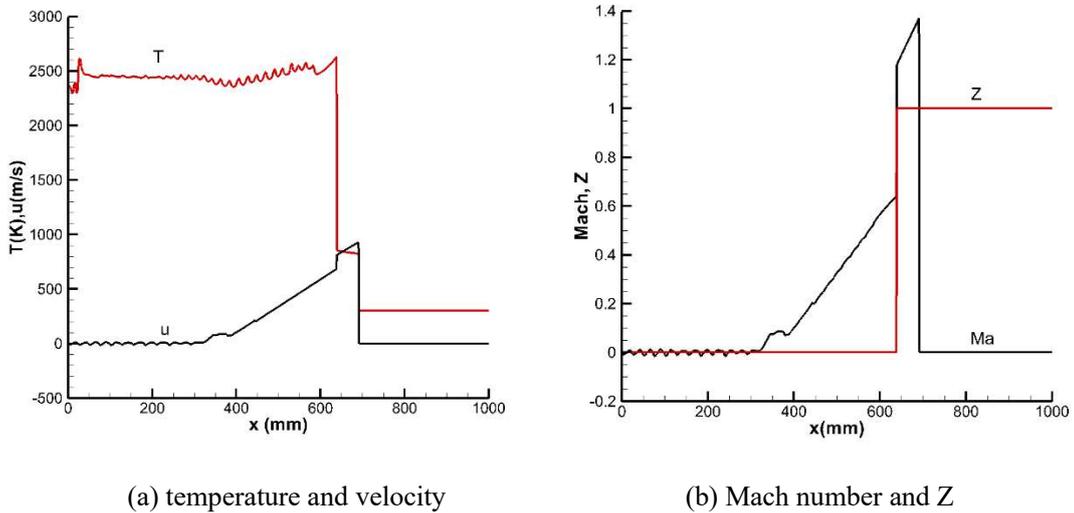

(a) temperature and velocity  (b) Mach number and Z

Fig.4 Profiles of different parameters of C-J deflagration

However, the total temperature of the flow induced by the leading shock wave is high enough to ignite the combustion. If the flow is slowed down by the rarefaction waves from the closed-end, the static temperature is increased, and the combustion is ignited again. But the combustion pressure is decreased by the rarefaction wave, which becomes a constant-pressure combustion. The propagating velocity of the rarefaction wave equals to the sound velocity of the combustion products. When the shock wave velocity decreases to the sound speed of the combustion products, the shock wave and the flame front will propagate with almost the same velocity and this double-discontinuity structure will keep quasi-steady for a certain distance. This is the essence of C-J deflagration.

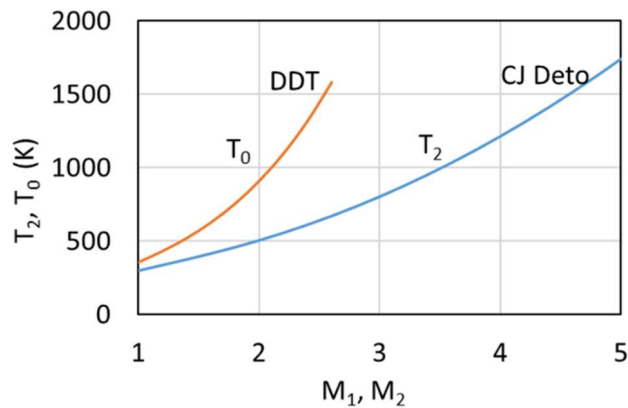

Fig.5 The total temperature induced by shock wave

The static temperate in the induction zone of C-J detonation and the total temperature induced by the C-J deflagration shock wave are plotted in Fig.5. We can find from Fig.5 that the total temperature induced by a shock wave with a Mach number of M2.5 is about 1500K, which is high enough to ignite the combustion. Note that this temperature is the total temperature of the induced flow, not the temperature of the reflected shock wave, which is higher than this value. In addition, we can give the key mechanism of DDT process here, which is illustrated in Fig.6. (The experimental results of [17] are used here to illustrate the DDT process.) If the Mach number is bigger than the DDT critical value, the total temperature is high enough to auto-ignite the mixture. When the flow is slowed down by reflection at the wall or by rarefaction waves, the DDT process occurs. If the Mach number is lower than the critical value, the total temperature is too low to ignite the mixture. Auto-ignition cannot occur, as a result, the DDT process cannot occur. Therefore, the essence of DDT process is from non-auto-ignition to auto-ignition.

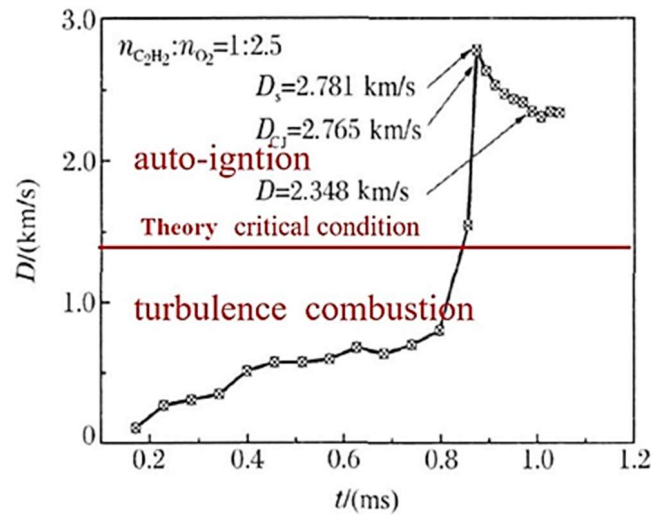

Fig.6 The essence of DDT process

**4 Conclusions**

In this study, the propagating mechanism of C-J deflagration is discussed. The C-J deflagration is created by decoupling the C-J detonation wave. Three different physical models are proposed to decouple the C-J detonation front. These three models are (a) to introduce an expansion parameter into the one-dimensional energy equation, (b) to increase the activation energy of the chemical reaction model and (c) to decouple the shock wave from the flame front by artificial method. One-dimensional numerical simulations are conducted and C-J deflagrations are obtained successfully by different models, chemical reaction kinetics and numerical schemes.

The essence of C-J deflagration is that the static temperature behind the leading shock wave is too low to ignite the mixture, but the total temperature of the induced flow is high enough to ignite the combustion. The induced flow is slowed down by rarefaction waves from the close-end and the constant-pressure combustion occurs. The flame front and the leading shock wave propagates with almost the same velocity as the sound speed of the combustion products, which is about 1/2 of the C-J detonation velocity. The essence of DDT process is from non-auto-ignition to auto-ignition.

**Acknowledgments**

This study is founded by the National Natural Science Foundation of China (No.11672312).